# X-ray warm absorber variability of the Seyfert Galaxy Arakelian 564

B. Korany [1,2]

And

M. I Nouh [1]

___________________________________________

[1]Department of Astronomy, National Research Institute of Astronomy and Geophysics ( NRISG), Cairo, Egypt
[2]Department of Physics, Faculty of Applied Science, Umm AL-Qura University, KSA

# Abstract

The aim of this paper is to study the variability of the warm absorber "clouds of ionised gas within AGN", for the Seyfert 1 galaxies Arakelian 564. The X-ray spectra for a four XMM-Newton observations of this object are analysed in EPIC and RGS instruments. These four observations covered a period of eleven years. There are some changes in the ionisation parameter ($\xi$) of the absorbing matter from one observation to another (log $\xi$ = 0.889±3.11×$10^{-2}$ for the year 2000 observation and 0.437±7.60×$10^{-2}$ for the year 2001). Also, the X-ray soft excess is studied for the four observation by using two black body parameters in EPIC spectra (the first black-body temperature is ~ 0.129±2.04×$10^{-3}$ Kev and the second is ~1.74 ±7.6 ×$10^{-2}$ Kev) and one black body parameter in RGS spectra.

Keywords galaxies: AGN X-rays: Warm absorber: Arakelian 564

# 1. Introduction

The first suggestion of the warm absorbers (absorption in the X-ray spectrum of AGN from ionized matter) is introduced by [1] to investigate Einstein data of the quasar MR 2251-178. We can note the warm absorber in the spectrum as a deficit of the soft-X-ray counts with respect to the power law energy distribution at energy ≈ 2Kev. The majority of the Seyfert galaxies contain ionized absorbing gas in their lines of sight [2], and there are various explanations for the origins of this warm absorber. One of these explanations is a wind formed by photoionised evaporation from the inner edge of the torus [3]. It is clear that, there is no definitive idea exists of the position of warm absorber in AGN. Some speculate is they could lie anywhere from the broad-line region to tens of parsecs out [3]. The warm absorber is also known to vary, and indeed the first discovered absorber was variable [4].

Arakelian 564 (Ark 564) [5,6,] is the X-ray brightest Narrow-line Seyfert 1 (NLS1) galaxy with a 2-10 Kev flux of ~ 2 × $10^{11}$ erg $cm^2 s^1$ and X-ray power law slope ($\Gamma$) = 2.43 ± 0.03, located at a redshift z = 0:02468. The presence of a soft excess was verified but neither the existence of a strong edge-like absorption feature at 0.712 keV nor the reported emission like feature at 1 keV could be confirmed [7]. Khanna et al. (2016) [8] found through long XMM-Newton RGS observations that the low ionization warm absorber is unusually low-velocity compared to other Seyferts. In addition to its broadband spectral features, Ark 564 has been studied with high-resolution spectrometers, and a low-ionization warm absorber is found [7].

We report in this paper on the variability of the warm absorber by studying four XMM-Newton observations (had chosen from 13 observations) cover a period of eleven years. We organized the paper as follows: the observations and data reduction in section 2, section 3 is devoted to the spectral analysis, while the results are summarised and discussed in section 4.

## 2. Observations and data reduction

The data used for the object had taken from Xmm-Newton archive. These data are observed in different four years (2000, 2001, 2005 and 2011). Table 1 contains the observations log of the data. All EPIC cameras were operated in the small-window mode and in the medium filter (except observation of 2011 in this filter). The data from the two EPIC MOS [10] cameras and the EPIC PN camera [11] are used. The raw data were processed with the EPIC pipeline chains of the Science Analysis System (SAS) software version 13.0.0. The event files of the individual observations have been cleaned from some bad time intervals characterised by high background events (so-called soft-proton fares). These bad time intervals rejected by creating light-curves for the observations (PN, MOS1, and MOS2), which are best visible above 10 keV. The spectrum of the object was extracted using circular extraction regions centered on the object, with a radius of 45 arc-sec for both the MOS and PN cameras. For the spectrum of the background, we used the same circular size from the source-free region [12]. All the spectra were subsequently analysed using the **xspec** software (v12.8.2).

By using RGSPROC task in XMMSAS the RGS data is reduced, and we extracted the spectra of the source and background with standard procedures. The first and second order spectra and response matrices from RGS1 and RGS2 were resampled to the first order RGS1 spectrum, then combined to produce a single spectrum and a single response matrix.

Table 1: XMM-Newton Observation log for ARK 564

| Year | Instr. | Mode | Filter | Exposure Start Time | Exposure End Time |
|---|---|---|---|---|---|
| 2000 | MOS1 | Small-window | MEDIUM FILTER | 2000-06-17@12:59:10 | 2000-06-17@16:12:19 |
| 2000 | MOS2 | Small-window | MEDIUM FILTER | 2000-06-17@12:56:5 | 2000-06-17@16:12:20 |
| 2000 | PN | Small-window | MEDIUM FILTER | 2000-06-17@12:08:54 | 2000-06-17@16:12:52 |
| 2001 | MOS1 | Small-window | MEDIUM FILTER | 2001-06-09@08:28:08 | 2001-06-09@11:54:45 |
| 2001 | MOS2 | Small-window | MEDIUM FILTER | 2001-06-09@08:30:14 | 2001-06-09@11:54:45 |
| 2001 | PN | Small-window | MEDIUM FILTER | 2001-06-09@08:38:08 | 2001-06-09@11:55:50 |
| 2005 | MOS1 | Small-window | MEDIUM FILTER | 2005-01-05@19:39:58 | 2005-01-06@23:17:22 |
| 2005 | MOS2 | Small-window | MEDIUM FILTER | 2005-01-05@19:39:59 | 2005-01-06@23:17:27 |
| 2005 | PN | Small-window | MEDIUM FILTER | 2005-01-05@19:47:37 | 2005-01-06@23:17:42 |
| 2011 | MOS1 | Small-window | THIN FILTER 1 | 2011-05-24@05:56:32 | 2011-05-24@22:27:09 |
| 2011 | MOS2 | Small-window | THIN FILTER 1 | 2011-05-24@05:56:32 | 2011-05-24@22:27:14 |
| 2011 | PN | Small-window | THIN FILTER 1 | 2011-05-24@06:02:05 | 2011-05-24@22:27:29 |

## 3. Spectral Analyses

To check if there is a warm absorber or not, a spectral analysis for the four observations of EPIC cameras (PN and MOS) in the broad band (0.4-10 keV) is done. The X-ray spectra were rebinned to contain at least 20 counts in each spectral bin using grppha command. All the data are fitted assuming a simple power law absorbed by the Galactic column density in the direction of the source which is fixed in the model with a constant value $5.24 \times 10^{20}$cm$^{-2}$ [13]. The simple absorbed power law model gives poor fit to the data (the best-fit models and residuals are shown in Figures (1 to 4)), providing a Reduced $\chi^2$/odf for example for PN data 3.2113/122 for first observation and 2.9809/111 for second observation (Table 2 list the Reduced $\chi^2$/odf for all the data). From the figures, it's clear that the fits are bad, but one can make out from the residuals a broad absorption trough between about 1 and 2 keV, and a soft excess under 1 kev. By comparing the spectra of the four observations, there is some difference in the soft part of the spectrum, suggesting some changes in the absorptions from one observation to another (Figures from 9 to 7).

### 3.1. Soft Excess

We started the treatment for the soft excess by using EPIC data of all observations. The hard band (2.5-10 Kev) data is fitted with a simple power-law model with the column of absorbing hydrogen atoms, as mentioned above, to avoid any soft excess. The fit is not bad, for the PN camera, the power law photon index ($\Gamma$) is $2.42 \pm 2.62 \times 10^{-2}$, $2.49 \pm 4.28 \times 10^{-2}$, $2.398 \pm 1.0367 \times 10^{-2}$ and $2.432 \pm 1.494 \times 10^{-2}$, and the reduced $\chi^2$/odf are 3.211/122, 2.981/111, 2.511/150 and 1.40/114 for 2000, 2001, 2005 and 2011 observations respectively, Table 3 summarises the fit parameters of all data. The average value for $\Gamma$ is $2,41 \pm 3.328 \times 10^{-2}$, which agrees with [13] and [14]. After that, we used zphabs (Multiplicative Model in xspec) component at the source redshift 0.024 [15] to test the presence of intrinsic cold absorption, and pcfabs (Multiplicative Model in xspec also) for testing the partial covering cold absorption. The fit statistics is not improved which mean the absence of a fully and partial covering intrinsic cold absorption. We didn't add the gaussian line model for FeK$\alpha$ emission line because it is not seen in the spectra, [14].

The fitted model (power law with absorbed column density) of the hard X-ray band (2.5 - 10.0 Kev) is extrapolated to the broadband spectral data (0.3 - 10 KeV), a huge soft X-ray excess found as shown in Figure 8.

We can provide a good fit to this soft excess by several models, such as single black bodies, multiple black bodies, multicolor dick black body, blurred reflection from partially ionized material, smeared absorption, and thermal computerization in the optically thick medium [16].

By using single black body model to fit the data after using power low model with absorption due to the hydrogen column density in our galaxy, (all parameters are free), the fit for all the observations is improved, with $\Delta\chi^2$ are 312, 210, 626 and 846 for observations of 2000, 2001, 2005 and 2011 respectively. The black-body temperature for all PN data are in between $0.129 \pm 3.02 \times 10^{-3}$ Kev to $0.160 \pm 4.1 \times 10^{-3}$ Kev or the average is 0.132 Kav. After adding another black body model, the fit is improved by $\Delta\chi^2$= 35.8; 37.7; 88.9 and 80.2 for observations of years 2000, 2001, 2005 and 2011 respectively, the first black-body temperature lies in the range $0.125\pm 3:02\times 10^{-3}$ to $01131\pm 2.2\times 10^{-3}$ Kev and the average value is $0.129 \pm 2.04 \times 10^{-3}$ and the second temperature lies in the range $1.59 \pm 9.9 \times 10^{-2}$ to $1.95 \pm 8:8 \times 10^{-2}$ Kev, with average value $1.74 \pm 7.6 \times 10^{-2}$ Kev.

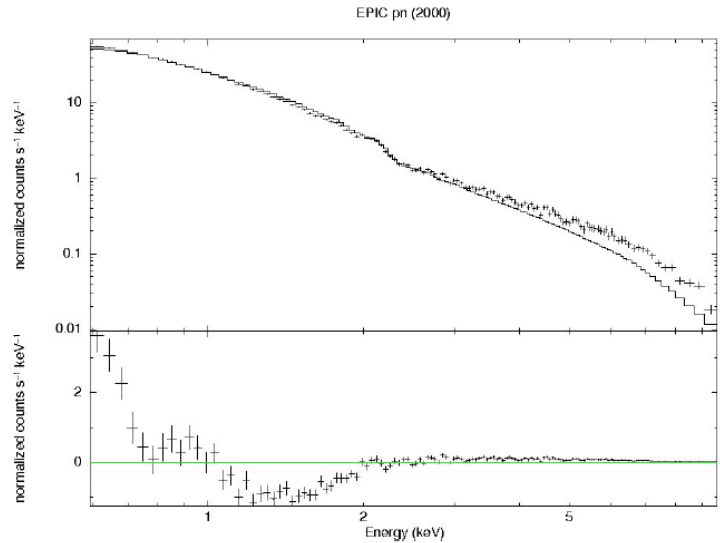

Fig. 1.— The fit of power law model with Nh freez for pn data of 2000

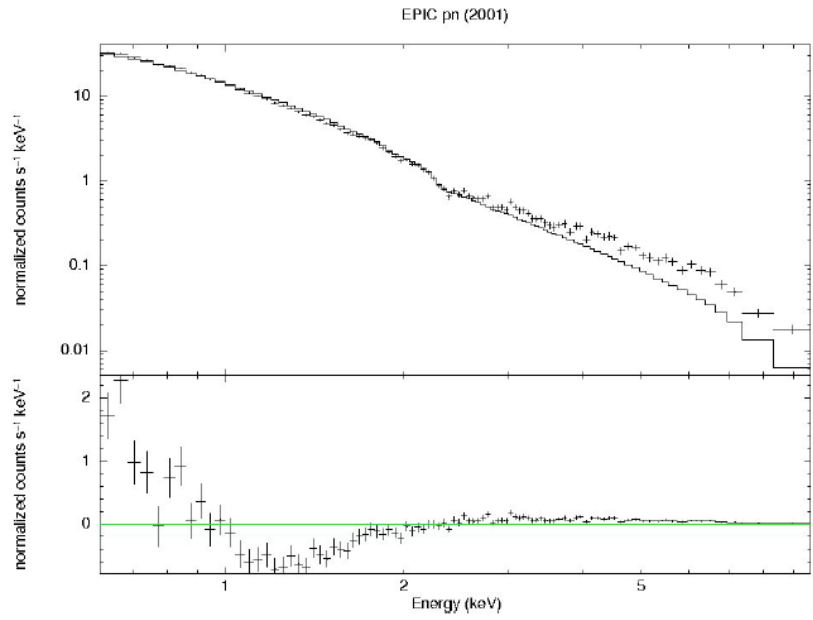

Fig. 2.— The fit of power law model with Nh freez for pn data of 2001

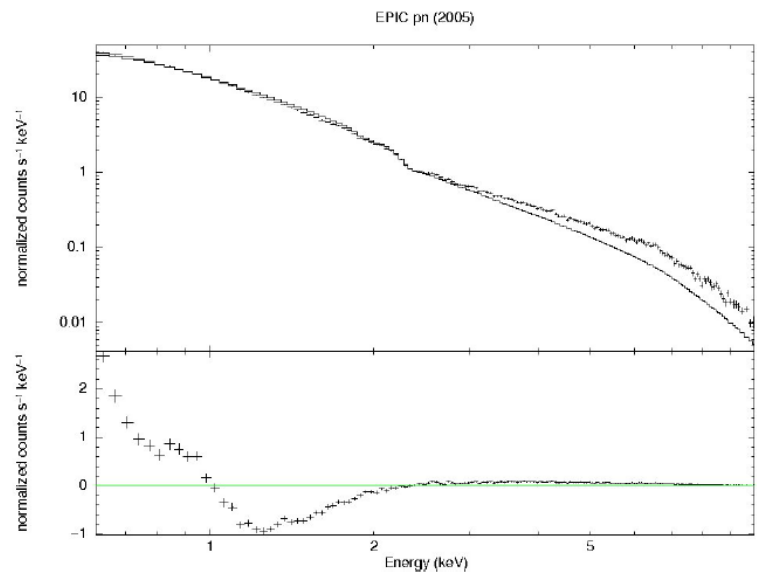

Fig. 3.— The fit of power law model with Nh freez for pn data of 2005

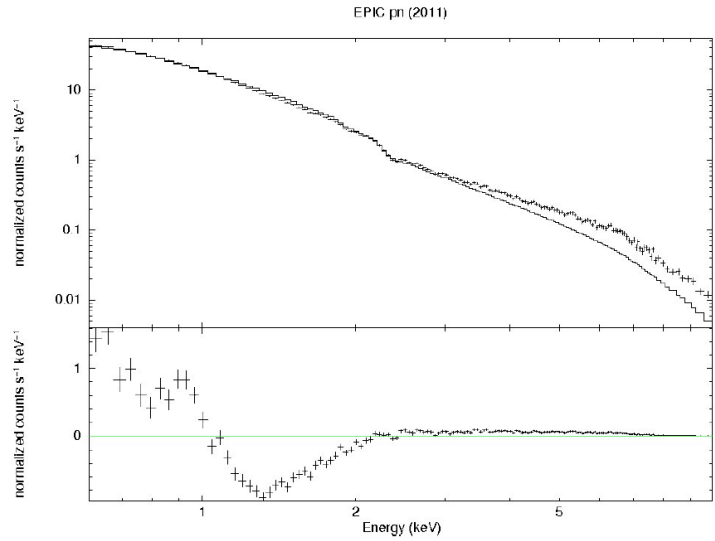

Fig. 4.— The fit of power law model with Nh freez for pn data of 2011

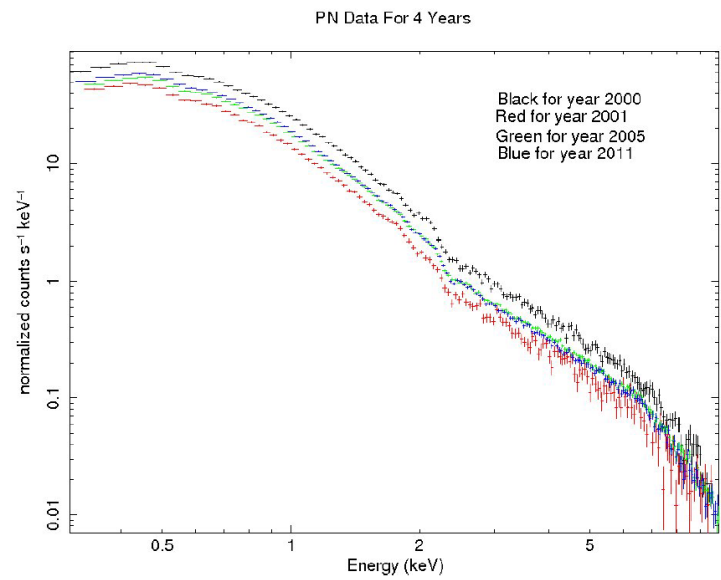

Fig. 5.— Four XMM-Newton data spectra shown together for comparison (PN)

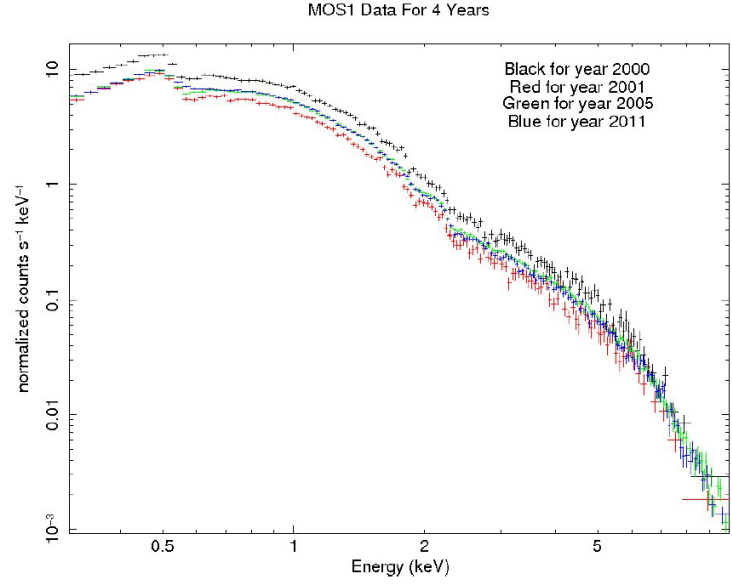

Fig. 6.— Four XMM-Newton data spectra shown together for comparison (MOS1)

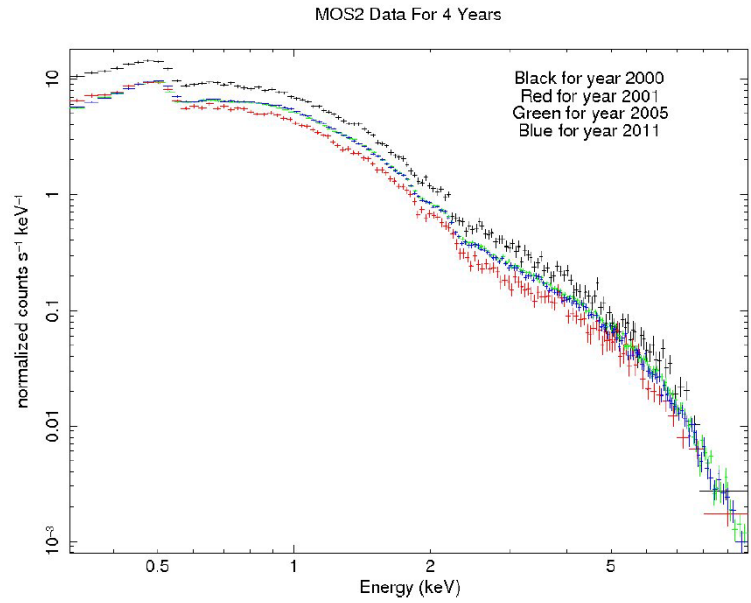

Fig. 7.— Four XMM-Newton data spectra shown together for comparison (MOS2)

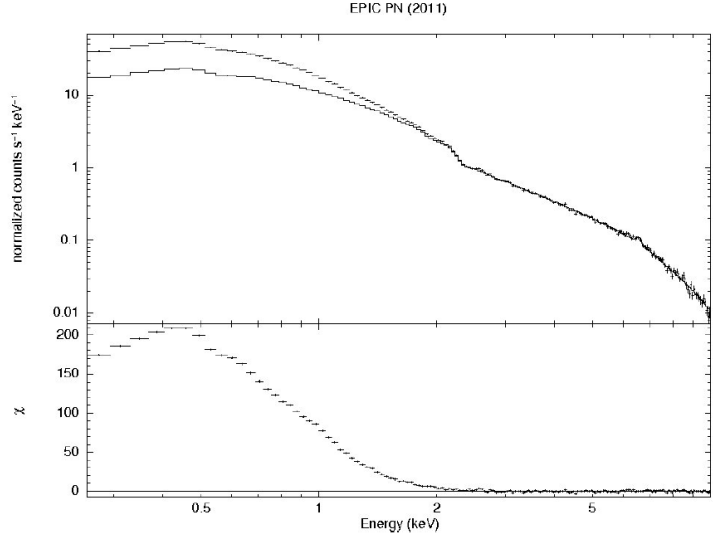

Fig. 8.— Extrapolated the fit of Hard to soft X-ray(Example for PN data of 2011)

Table 2: The Fit Prametars for Pl with Nh for all data

| Year | Instroment | Reduced $\chi^2/odf$ |
|---|---|---|
| 2000 | MOS1 | 9.51407 for 144 |
| 2000 | MOS2 | 3.1984 for 116 |
| 2000 | PN | 3.2113 for 122 |
| 2001 | MOS1 | 6.8267 for 139 |
| 2001 | MOS2 | 3.6585 for 109 |
| 2001 | PN | 2.9809 for 111 |
| 2005 | MOS1 | 6.1554 for 164 |
| 2005 | MOS2 | 3.097 for 145 |
| 2005 | PN | 2.5113 for 150 |
| 2011 | MOS1 | 3.1328 for 162 |
| 2011 | MOS2 | 16.2792 for 141 |
| 2011 | PN | 13.9961 for 146 |

Table 3: Th fit Prametars for PL with Nh for all data in the hard band (2.5 -10.0 Kev)

| Year | Instrument | Γ | Normalization | Reduced $\chi^2/odf$ |
|---|---|---|---|---|
| 2000 | MOS1 | $2.37 \pm 5.14 \times 10^{-2}$ | $1.49 \times 10^{-2} \pm 1.03 \times 10^{-3}$ | 1.02 for 63 |
| 2000 | MOS2 | $2.54 \pm 5.05 \times 10^{-2}$ | $1.9 \times 10^{-2} \pm 1.30 \times 10^{-3}$ | 1.25 for 67 |
| 2000 | PN | $2.42 \pm 2.62 \times 10^{-2}$ | $1.65 \times 10^{-2} \pm 6.037 \times 10^{-4}$ | 3.211 for 122 |
| 2001 | MOS1 | $2.39 \pm 6.29 \times 10^{-2}$ | $9.09 \times 10^{-3} \pm 7.67 \times 10^{-4}$ | 6.827 for 139 |
| 2001 | MOS2 | $2.34 \pm 6.29 \times 10^{-2}$ | $7.94 \times 10^{-3} \pm 6.768 \times 10^{-4}$ | 3.659 for 109 |
| 2001 | PN | $2.49 \pm 4.28 \times 10^{-2}$ | $8.72 \times 10^{-3} \pm 5.050 \times 10^{-4}$ | 2.981 for 111 |
| 2005 | MOS1 | $2.35 \pm 1.60 \times 10^{-2}$ | $1.12 \times 10^{-2} \pm 2.418 \times 10^{-4}$ | 6.156 for 164 |
| 2005 | MOS2 | $2.39 \pm 1.61 \times 10^{-2}$ | $1.164 \times 10^{-2} \pm 2.533 \times 10^{-4}$ | 1.52 for 116 |
| 2005 | PN | $2.398 \pm 1.0367 \times 10^{-2}$ | $1.141 \times 10^{-2} \pm 1.66 \times 10^{-4}$ | 2.511 for 150 |
| 2011 | MOS1 | $2.377 \pm 2.282 \times 10^{-2}$ | $1.081 \times 10^{-2} \pm 3.320 \times 10^{-4}$ | 1.51 for 94 |
| 2011 | MOS2 | $2.430 \pm 2.243 \times 10^{-2}$ | $1.164 \times 10^{-2} \pm 3.517 \times 10^{-4}$ | 1.10 for 98 |
| 2011 | PN | $2.432 \pm 1.494 \times 10^{-2}$ | $1.124 \times 10^{-2} \pm 2.320 \times 10^{-4}$ | 1.40 for 114 |

Table 4: Results for warm absorber( PN data)

| Y | M* | Nh(Pha) | Γ | $bb_1$ | $bb_2$ | Nh(zxipcf) | log ξ | $\chi^2/odf$ |
|---|---|---|---|---|---|---|---|---|
| 2000 | M1 | $5.22 \times 10^{20}$ $\pm 9.0 \times 10^{-3}$ | 2.89 $\pm 7.7 \times 10^{-3}$ | .... .... | .... .... | .... ... | .... .... | 1674/152 |
| 2000 | M2 | $3.7 \times 10^{20}$ $\pm 1.6 \times 10^{-3}$ | 2.55 $\pm 1.2 \times 10^{-2}$ | 0.133 $\pm 1.25 \times 10^{-3}$ | .... .... | .... .... | .... .... | 324/150 |
| 2000 | M3 | $5.6 \times 10^{20}$ $\pm 2.56 \times 10^{-3}$ | 2.84 $\pm 3.2 \times 10^{-3}$ | 0.130 $\pm 2.2 \times 10^{-3}$ | 1.7 $\pm 8.7 \times 10^{-2}$ | .... ... | .... ... | 218/148 |
| 2000 | M4 | $6.9 \times 10^{20}$ $\pm 6.6 \times 10^{-3}$ | 3.08 $\pm 1.2 \times 10^{-1}$ | 0.158 $\pm 1.4 \times 10^{-3}$ | 2.02 $\pm 2.0 \times 10^{-1}$ | $2.34 \times 10^{22}$ $\pm 3.9 \times 10^{-1}$ | 0.889 $\pm 3.11 \times 10^{-2}$ | 140/145 |
| 2001 | M1 | $4.27 \times 10^{20}$ $\pm 1.55 \times 10^{-3}$ | 2.99 $\pm 1.21 \times 10^{-2}$ | .... .... | .... .... | .... ... | .... .... | 1001/146 |
| 2001 | M2 | $2.44 \times 10^{20}$ $\pm 2.55 \times 10^{-3}$ | 2.59 $\pm 1.8 \times 10^{-2}$ | 0.129 $\pm 2.04 \times 10^{-3}$ | .... .... | .... .... | .... .... | 255/144 |
| 2001 | M3 | $5.08 \times 10^{20}$ $\pm 3.9 \times 10^{-3}$ | 2.9 $\pm 4.7 \times 10^{-2}$ | 0.125 $\pm 3.02 \times 10^{-3}$ | 1.59 $\pm 9.9 \times 10^{-2}$ | .... ... | .... ... | 166/142 |
| 2001 | M4 | $7.12 \times 10^{20}$ $\pm 6.7 \times 10^{-3}$ | 3.07 $\pm 9.5 \times 10^{-2}$ | 0.160 $\pm 4.1 \times 10^{-3}$ | 1.90 $\pm 1.7 \times 10^{-1}$ | $2.7 \times 10^{22}$ $\pm 3.9 \times 10^{-1}$ | 0.437 $\pm 7.60 \times 10^{-2}$ | 140/139 |
| 2005 | M1 | $5.37 \times 10^{20}$ $\pm 3.66 \times 10^{-4}$ | 2.93 $\pm 2.37 \times 10^{-3}$ | .... .... | .... .... | .... ... | .... .... | 12394/172 |
| 2005 | M2 | $3.9 \times 10^{20}$ $\pm 5.75 \times 10^{-4}$ | 2.55 $\pm 1.2 \times 10^{-2}$ | 0.130 $\pm 5.1 \times 10^{-4}$ | .... .... | .... .... | .... .... | 1482.2/170 |
| 2005 | M3 | $5.78 \times 10^{20}$ $\pm 9.10 \times 10^{-4}$ | 2.86 $\pm 1.2 \times 10^{-2}$ | 0.130 $\pm 7.2 \times 10^{-4}$ | 1.7 $\pm 3.0 \times 10^{-2}$ | .... ... | .... ... | 632/168 |
| 2005 | M4 | $3.7 \times 10^{20}$ $\pm 2.3 \times 10^{-3}$ | 2.71 $\pm 1.85 \times 10^{-3}$ | 0.137 $\pm 1.7 \times 10^{-3}$ | 1.8 $\pm 6.5 \times 10^{-2}$ | $2.34 \times 10^{22}$ $\pm 3.9 \times 10^{-1}$ | 0.578 $\pm 1.07 \times 10^{-2}$ | 301/160 |
| 2011 | M1 | $7.28 \times 10^{20}$ $\pm 5.58 \times 10^{-2}$ | 3.01 $\pm 4.33 \times 10^{-3}$ | .... .... | .... .... | .... ... | .... .... | 8536/169 |
| 2011 | M2 | $4.68 \times 10^{20}$ $\pm 9.0 \times 10^{-4}$ | 2.59 $\pm 6.56 \times 10^{-3}$ | 0.135 $\pm 6.7 \times 10^{-4}$ | .... .... | .... .... | .... .... | 767/167 |
| 2011 | M3 | $6.2 \times 10^{20}$ $\pm 1.36 \times 10^{-3}$ | 2.80 $\pm 1.66 \times 10^{-2}$ | 0.131 $\pm 2.2 \times 10^{-3}$ | 1.95 $\pm 8.8 \times 10^{-2}$ | .... ... | .... ... | 487/165 |
| 2011 | M4 | $5.78 \times 10^{20}$ $\pm 6.6 \times 10^{-3}$ | 2.78 $\pm 2.03 \times 10^{-2}$ | 0.136 $\pm 1.3 \times 10^{-3}$ | 2.11 $\pm 1.4 \times 10^{-1}$ | $1.7 \times 10^{22}$ $\pm 6.9 \times 10^{-2}$ | 0.653 $\pm 8.7 \times 10^{-2}$ | 221/163 |

* M means the model name, where M1= Pha+PL(powr low), M2= Pha+PL+ bb(black body model),
M3= Pha+PL+bb+bb, and M4=Pha+PL+bb+bb+zxipcf(partial covering absorption by partially ionized material model.

Table 5: Results for warm absorber( RGS data)

| Y | M | Nh(Pha) | $\Gamma$ | $bb_1$ | Nh(zxipcf) | $\log \xi$ | $\chi^2/odf$ |
|---|---|---|---|---|---|---|---|
| 2000 | M1 | $8.97 \times 10^{20}$ | 3.24 | .... | .... | .... | 2911/2457 |
|  |  | $\pm 3.4 \times 10^{-3}$ | $\pm 3.9 \times 10^{-2}$ | .... | ... | .... |  |
|  | M2 | $6.56 \times 10^{20}$ | 2.89 | 0.127 | .... | .... | 2756/2455 |
|  |  | $\pm 4.04 \times 10^{-3}$ | $\pm 5.04 \times 10^{-2}$ | $\pm 3.75 \times 10^{-3}$ | .... | .... |  |
|  | M3 | $6.13 \times 10^{20}$ | 2.82 | 0.131 | $6.39 \times 10^{20}$ | 0.800 | 2697/2452 |
|  |  | $\pm 4.22 \times 10^{-3}$ | $\pm 5.48 \times 10^{-2}$ | $\pm 3.26 \times 10^{-3}$ | $\pm 2.24 \times 10^{-2}$ | $\pm 1.15 \times 10^{-1}$ |  |
| 2001 | M1 | $8.92 \times 10^{20}$ | 3.36 | .... | .... | .... | 2706/2446 |
|  |  | $\pm 5.98 \times 10^{-3}$ | $\pm 5.15 \times 10^{-2}$ | .... | ... | .... |  |
|  | M2 | $6.50 \times 10^{20}$ | 3.10 | 0.134 | .... | .... | 2650/2444 |
|  |  | $\pm 6.93 \times 10^{-3}$ | $\pm 8.5 \times 10^{-2}$ | $\pm 9.22 \times 10^{-3}$ | .... | .... |  |
|  | M3 | $4.71 \times 10^{20}$ | 2.98 | 0.131 | $0.14 \times 10^{22}$ | 0.434 | 2633/2441 |
|  |  | $\pm 8.40 \times 10^{-3}$ | $\pm 1.10 \times 10^{-1}$ | $\pm 5.15 \times 10^{-3}$ | $\pm 7.99 \times 10^{-2}$ | $\pm 9.20 \times 10^{-2}$ |  |
| 2005 | M1 | $7.74 \times 10^{20}$ | 3.20 | .... | .... | .... | 3878/2445 |
|  |  | $\pm 2.01 \times 10^{-3}$ | $\pm 1.66 \times 10^{-2}$ | .... | ... | .... |  |
|  | M2 | $5.14 \times 10^{20}$ | 2.76 | 0.123 | .... | .... | 3340/2443 |
|  |  | $\pm 2.38 \times 10^{-4}$ | $\pm 3.16 \times 10^{-2}$ | $\pm 1.75 \times 10^{-3}$ | .... | .... |  |
|  | M3 | $4.58 \times 10^{20}$ | 2.68 | 0.130 | $0.10 \times 10^{22}$ | 0.562 | 3167/2440 |
|  |  | $\pm 2.49 \times 10^{-3}$ | $\pm 3.45 \times 10^{-2}$ | $\pm 1.68 \times 10^{-3}$ | $\pm 2.84 \times 10^{-2}$ | $\pm 6.07 \times 10^{-2}$ |  |
| 2011 | M1 | $9.28 \times 10^{20}$ | 3.31 | .... | .... | .... | 3317/2448 |
|  |  | $\pm 2.86 \times 10^{-3}$ | $\pm 2.33 \times 10^{-2}$ | .... | ... | .... |  |
|  | M2 | $5.82 \times 10^{20}$ | 2.78 | 0.127 | .... | .... | 2867/2446 |
|  |  | $\pm 3.41 \times 10^{-3}$ | $\pm 4.45 \times 10^{-2}$ | $\pm 2.25 \times 10^{-3}$ | .... | .... |  |
|  | M3 | $5.34 \times 10^{20}$ | 2.71 | 0.133 | $.15 \times 10^{22}$ | 0.607 | 2780/2443 |
|  |  | $\pm 3.51 \times 10^{-3}$ | $\pm 5.02 \times 10^{-2}$ | $\pm 1.3 \times 10^{-3}$ | $\pm 5.24 \times 10^{-2}$ | $\pm 8.02 \times 10^{-2}$ |  |

## 3.2. Warm Absorber

Now we begin to study the warm absorber in the broad band (0.4- 10 keV) spectra for EPIC data. In addition, the absorbed power law model and the models used to study the soft excess, we added a warm absorber phase, which could be defined as a gas at a particular ionisation parameter($\xi$) and column density. The ionisation parameter writes as $\xi = L/nr^2$ erg cm s$^{-1}$, where L is the ionising luminosity, n is the gas density, and r the distance of the ionising source from the absorbing gas, [17]. This warm absorber was modeled with the partial covering absorption by partially ionized material (zxipcf in xspec), which uses a grid of XSTAR photoionised absorption models for the absorption, then assumes that this only covers some fraction f of the source, while the remaining (1-f) of the spectrum is seen directly. The micro-turbulent velocity is assumed to be 200 km/s. The proposed model is added to all free parameters except the redshift component. The redshift is frozen at 0.024 [15], an improvement to the fit is found ($\Delta\chi^2$ are 27, 9,22, and 98 for observations 2000, 2001, 2005 and 2011 respectively), Fig 9 plots the fit of the all parameters.

From the fitting we can see after adding the warm absorber model, there is no change occur in the parameters except a small change in the black body temperatures and in the photon index ($\Gamma$) of the power law model for data of years 2000 and 2001. The ($\Gamma$) changed from 2.84 to 3.08 and from 2.9 to 3.07. The temperatures changed from 0.13 and 1.7 Kev to 0.158 and 2.02 Kev for observation of year 2000, and form 0.125 and 1.5 Kev to 0.16 and 1.9 Kev for observation of the year 2001. The best fit values of the parameters of the partial covering absorption by partially ionized material model are, the column density ($Nh_{wa}$) are 2.34 ×10$^{22}$ ±3.9 ×10$^{-1}$; 2.7 × 10$^{22}$ ±3.9 ×10$^{-1}$; 2.34 ×10$^{22}$ ±3.9 ×10$^{-1}$and 1.7 ×10$^{22}$ ±6.9 ×10$^{-1}$ for the observations of the years 2000, 2001, 2005 and 2011 respectively. From table 4 we can see from the values of the photonionised absorption parameter (log $\xi$) the evidence of the warm absorber in the range 1.0 - 2.0 Kav range. The values of log $\xi$ are 0.889 ± 3.11 ×10$^{-2}$; 0.437 ± 7.60 ×10$^{-2}$; 0.578 ±1.07 ×10$^{-2}$; and 0.653±8.7×10$^{-2}$respectively (see table 4). The different values of the log $\xi$ parameter from one observation to another means that there are some changes in the warm absorber from one data to another, and we can see that the highest value is for log ξ (0.889 ± 3.11 × $10^{-2}$) from the observation of year 2000, while the least value (0.437 ± 7.60 ×$10^{-2}$) from observation of the year 2001.

Because of the presence of warm absorption features may become more evident in RGS spectrum, the RGS data in the range 0.4 to 2.0 Kev is used with a model similar to that used with EPIC data i.e., absorbed power law model with two black body parameters and one warm absorber model. But, by adding the first black body model to the absorbed power law model, the temperatures are 0.127_3.75×$10^{-3}$; 0.134_9.22 × $10^{-3}$; 0.123×1.75×$10^{-3}$ and 0.127 × 2.25 ×$10^{-3}$. The average value for these temperatures is approximately the same value as the temperature of the EPEC data. After adding the second black body the fit is not improved. In this model one black body parameter is sufficient. Table 5 summaries the fitting output, we can see from the results that the values of the photoionization parameter are approximately the same value as the EPIC data, Fig 10 illustrates the fitting after adding all models.

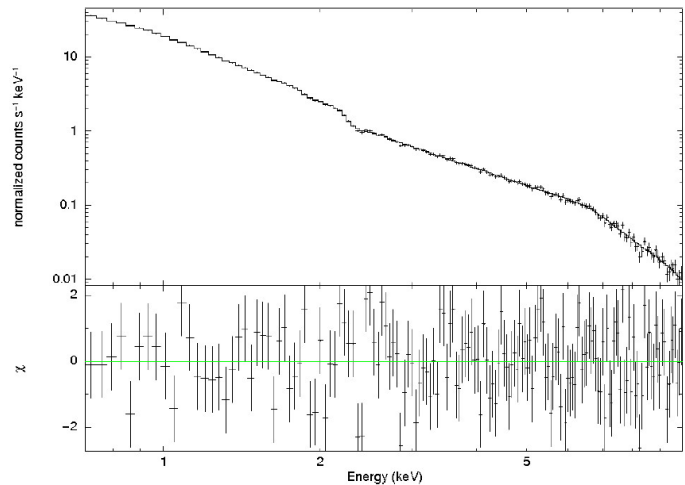

Fig. 9.— The final fit of the models after adding the warm absorber model (Example for PN data of 2011)

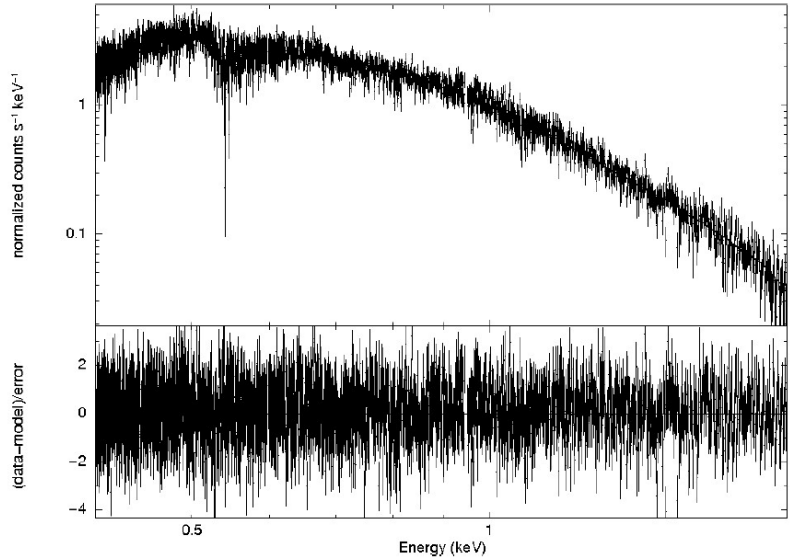

Fig. 10.— The final fit of the models after adding the warm absorber model for RGS data(observation of 2011 as example )

## 4. Conclusions

A detailed spectral analysis (EPIC and RGS) have been performed for four XMM-Newton observations to the Seyfert galaxy Arakelian 564 (observations of years 2000, 2001, 2005 and 2011). The wide range (0.35-10.0 Kev) is described by a power law model with a photon index $\Gamma \_ 2:41$, and the soft excess consists of two black body components in the EPIC data with temperatures ~ 0.129 Kev and 1.74 Kev, and one black body component in RGS data. The test to the presence of intrinsic cold absorption, gives us that there is no fully and partial covering intrinsic cold absorption. We did not add the Gaussian line model for Fe Kα emission line because it is not seen in the spectra. The spectra were almost constant for the four observations at the hard part which is > 5 Kev, but the soft part changed from one observation to another. We detected one-phase warm absorber in the four observation spectra by using the partial covering absorption using a partially ionized material model. The value of the photoionisation parameter changed from one spectrum to another. The highest value is for log ξ(0.889 ± 3.11×10$^{-2}$) from the observation of the year 2000, while the least value is from observation of the year 2001 for the EPIC. The output results from the RGS data are approximately match with EPIC data. A long-time observation may be used in the near future to shed more light on the variability of the warm absorber.